% modifs since submitted:
% misprints, refs updated
% more data in table and fits refined
%
% redo all the fits when p=22 obtained

\input harvmac
\input labeldefs.tmp
\writedefs
\overfullrule=0pt

% redefine figures ... 
\input epsf
\def\fig#1#2#3{
\xdef#1{\the\figno}
\writedef{#1\leftbracket \the\figno}
\nobreak
\par\begingroup\parindent=0pt\leftskip=1cm\rightskip=1cm\parindent=0pt
\baselineskip=11pt
\midinsert
\centerline{#3}
\vskip 12pt
{\bf Fig.\ \the\figno:} #2\par
\endinsert\endgroup\par
\goodbreak
\global\advance\figno by1
}
\newwrite\tfile\global\newcount\tabno \global\tabno=1
\def\tab#1#2#3{
\xdef#1{\the\tabno}
\writedef{#1\leftbracket \the\tabno}
\nobreak
\par\begingroup\parindent=0pt\leftskip=1cm\rightskip=1cm\parindent=0pt
\baselineskip=11pt
\midinsert
\centerline{#3}
\vskip 12pt
{\bf Tab.\ \the\tabno:} #2\par
\endinsert\endgroup\par
\goodbreak
\global\advance\tabno by1
}
\def\der{\partial}

\font\cmss=cmss10 \font\cmsss=cmss10 at 7pt
\def\R{\relax{\rm I\kern-.18em R}}
\def\Z{\relax\ifmmode\mathchoice
{\hbox{\cmss Z\kern-.4em Z}}{\hbox{\cmss Z\kern-.4em Z}}
{\lower.9pt\hbox{\cmsss Z\kern-.4em Z}}
{\lower1.2pt\hbox{\cmsss Z\kern-.4em Z}}\else{\cmss Z\kern-.4em Z}\fi}
\def\bra#1{\big< #1 \big|\,}
\def\ket#1{\,\big| #1 \big>}
\def\braket#1#2{\big< #1 \big| #2 \big>}
\def\Th{Thistlethwaite}
\def\Gam{{\mit\Gamma}}
%
%%%%%%%%%%%%%%%%%%%%%%%
% preprint # in refs ?
\def\pre#1{ (preprint {\tt #1})}%use this to give preprint # in refs
%\def\pre#1{}%use this NOT to give preprint # in refs
%%%%%%%%%%%%%%%%%%%%%%%%%%%%%%%%%%%%%%%%%%%%%%%%%%%%%%%%%%%%
%
%References
%
%\lref\XXX{N.N. XXX, {\it title }, .}
% \refs{\..{--}\...}
%
% update some refs
%
\lref\TutteMap{W.~T.~Tutte,
{\sl A census of planar maps},
{\it Canad.~J.~Math.} {\bf 15}, 249--271 (1963).}
\lref\TutteHam{W.~T.~Tutte,
{\sl A census of Hamiltonian polygons},
{\it Canad.~J.~Math.} {\bf 14} 402-417 (1962).}
\lref\KPZ{V.~G.~Knizhnik, A.~M.~Polyakov and A.~B.~Zamolodchikov,
{\sl Fractal structure of 2D quantum gravity},
{\it Mod.~Phys.~Lett.~A} {\bf 3}, 819--826 (1988);
F.~David,
{\sl Conformal field theories coupled to 2D gravity in the conformal gauge},
{\it Mod.~Phys.~Lett.~A} {\bf 3}, 1651--1656 (1988);
J.~Distler and H.~Kawai,
{\sl Conformal field theory and 2D quantum gravity},
{\it Nucl.~Phys.} {\bf B 321}, 509 (1989).}
\lref\Hash{R.~Sedgewick,
{\sl Algorithms in C} (Addison-Wesley, 1990).}
\lref\gusein{S.~M.~Gusein-Zade,
{\it Adv.~Soviet Math.} {\bf 21}, 189--198 (1994);
S.~M.~Gusein-Zade and F.~S.~Duzhin,
{\sl On the number of topological types of plane curves},
{\it Uspekhi Math.~Nauk.} {\bf 53}, 197--198 (1998)
[English translation in {\it Russian Math.~Surveys} {\bf 53}, 626--627 (1998)].}
\lref\Knuth{D.~E.~Knuth,
{\sl The Art of Computer Programming}, vol 2: {\sl Seminumerical Algorithms}
(Addison-Wesley, 1969).}
\lref\Dyck{M.-P.~Delest and G.~Viennot,
{\it Theor.~Comp.~Sci.} {\bf 34} (1984) 169.}
\nref\Lorentz{P.~Di Francesco, E.~Guitter and C.~Kristjansen,
{\sl Integrable 2D Lorentzian Gravity and Random Walks},
{\it Nucl.~Phys.} {\bf B 567} (2000) 515--553\pre{hep-th/9907084}.}
\nref\Jensen{I.~Jensen,
{\sl Enumerations of Plane Meanders} \pre{cond-mat/9910313};
{\sl A Transfer Matrix Approach to the Enumeration of Plane Meanders},
{\it J.~Phys.~A}, to appear\pre{cond-mat/0008178}.}
\nref\Meanders{P.~Di Francesco, E.~Guitter and J.L.~Jacobsen,
{\sl Exact Meander Asymptotics: a Numerical Check},
{\it Nucl.~Phys.} {\bf B 580} (2000) 757--795
\pre{cond-mat/0003008}.}
\nref\Ising{P.~Di Francesco, E.~Guitter and J.L.~Jacobsen,
work in progress.}
\lref\AV{I.Ya.~Arefeva and I.V.~Volovich, 
{\sl Knots and Matrix Models}, {\it Infinite Dim.
Anal.~Quantum Prob.} {\bf 1} (1998) 1\pre{hep-th/9706146}).}
\lref\BIPZ{E.~Br{\'e}zin, C.~Itzykson, G.~Parisi and J.-B.~Zuber, 
{\sl Planar Diagrams}, {\it Commun.~Math.~Phys.} {\bf 59} (1978) 35--51.}
\lref\BIZ{D.~Bessis, C.~Itzykson and J.-B.~Zuber, 
{\sl Quantum Field Theory Techniques in Graphical Enumeration},
{\it Adv.~Appl.~Math.} {\bf 1} (1980) 109--157.}
\lref\DFGZJ{P.~Di Francesco, P.~Ginsparg and J.~Zinn-Justin, 
{\sl 2D Gravity and Random Matrices, }{\it Phys.~Rep.} {\bf 254} (1995)
1--133.}
\lref\tH{G.~'t Hooft, 
{\sl A Planar Diagram Theory for Strong 
Interactions}, {\it Nucl.~Phys.} {\bf B 72} (1974) 461--473.}
\lref\HTW{J.~Hoste, M.~Thistlethwaite and J.~Weeks, 
{\sl The First 1,701,936 Knots}, {\it The Mathematical Intelligencer}
{\bf 20} (1998) 33--48.}
\lref\MTh{W.W.~Menasco and M.B.~\Th, 
{\sl The Tait Flyping Conjecture}, {\it Bull.~Amer.~Math.~Soc.} {\bf 25}
(1991) 403--412; 
{\sl The Classification of Alternating 
Links}, {\it Ann.~Math.} {\bf 138} (1993) 113--171.}
\lref\Ro{D.~Rolfsen, {\sl Knots and Links}, Publish or Perish, Berkeley 1976.}
\lref\STh{C.~Sundberg and M.~Thistlethwaite, 
{\sl The rate of Growth of the Number of Prime Alternating Links and 
Tangles}, {\it Pac.~J.~Math.} {\bf 182} (1998) 329--358.}
\lref\Tutte{W.T.~Tutte, {\sl A Census of Planar Maps}, 
{\it Can.~J.~Math.} {\bf 15} (1963) 249--271.}
\lref\Zv{A.~Zvonkin, {\sl Matrix Integrals and Map Enumeration: An Accessible
Introduction},
{\it Math.~Comp.~Modelling} {\bf 26} (1997) 281--304.}
\lref\KM{V.A.~Kazakov and A.A.~Migdal, {\sl Recent progress in the
theory of non-critical strings}, {\it Nucl.~Phys.} {\bf B 311} (1988)
171--190.}
\lref\KP{V.A.~Kazakov and P.~Zinn-Justin, {\sl Two-Matrix Model with
$ABAB$ Interaction}, {\it Nucl.~Phys.} {\bf B 546} (1999) 647\pre{hep-th/9808043}.}
\nref\ZJZ{P.~Zinn-Justin and J.-B.~Zuber, {\sl Matrix Integrals
and the Counting of Tangles and Links},
to appear in the proceedings of the 11th 
International Conference on Formal Power Series and Algebraic 
Combinatorics, Barcelona June 1999
\pre{math-ph/9904019}.}
\nref\PZJ{P.~Zinn-Justin, {\sl Some Matrix Integrals
related to Knots and Links}, proceedings
of the 1999 semester of the MSRI ``Random Matrices
and their Applications'', MSRI Publications Vol. 40 (2001)
\pre{math-ph/9910010}.}
\nref\ZJZb{P.~Zinn-Justin and J.-B.~Zuber, {\sl On the Counting of Colored
Tangles}, {\it Journal of Knot Theory and its Ramifications} 
9 (2000) 1127--1141\pre{math-ph/0002020}.}
\lref\PZJb{P.~Zinn-Justin, {\sl The Six-Vertex Model on
Random Lattices}, {\it Europhys.~Lett.} 50 (2000) 15--21\pre{cond-mat/9909250}.}
\lref\IK{I.~Kostov, {\sl Exact solution of the Six-Vertex
Model on a Random Lattice}, {\it Nucl.~Phys.} {\bf B 575} (2000) 513-534\pre{hep-th/9911023}.}
\nref\Nech{S.~Nechaev, {\sl Statistics of knots and entangled random walks},
lectures at Les Houches 1998 summer school\pre{cond-mat/9812205}.}
\nref\KAUF{L.H.~Kauffman, {\sl Knots and physics},
World Scientific Pub Co (1994).}
\lref\JZJ{J.~L.~Jacobsen and P.~Zinn-Justin, {\sl A Transfer Matrix approach to the Enumeration of Colored Links}\pre{math-ph/0104009}.}
\lref\PZJc{P.~Zinn-Justin, {\sl The General O(n) Quartic Matrix Model and its application to Counting Tangles and Links}\pre{math-ph/0106005}.}
%
%%%%%%%%%%%%%%%%%%%%%%%%%%%%%%%%%%%%%%%%%%%%%%%%%%%%%%%%%%%%%%%%%%%%%
\Title{\tt math-ph/0102015}
{{\vbox {
\vskip-10mm
\centerline{A Transfer Matrix approach}
\vskip2pt
\centerline{to the Enumeration of Knots}
}}}
\medskip
\centerline{J.~L.~Jacobsen {\it and} P.~Zinn-Justin}\medskip
\centerline{\sl Laboratoire de Physique Th\'eorique et Mod\`eles Statistiques}
\centerline{\sl Universit\'e Paris-Sud, B\^atiment 100}
\centerline{\sl 91405 Orsay Cedex, France}
\vskip .2in
% abstract
\noindent 

We propose a new method to enumerate alternating knots using a
transfer matrix approach. We apply it to count numerically various
objects, including prime alternating
tangles with two connected components, up to order $18$--$22$, and comment
on the large-order behavior in connection with one of the authors' conjecture.

\Date{02/2001}
%\draft

\newsec{Introduction}
The classification of knots is a subject with a long history.
More than a hundred years ago, Tait, Kirkman and Little tried to draw
``by hand'' the first knots (up to 10 crossings, but with some mistakes);
nowadays, it is possible
with the modern tools of knot theory and the use of computers to create
programs that generate all possible (prime) knots up to 16 crossings
\HTW. Here we shall consider the simplest problem,
namely the {\it enumeration} of knots (or similar objects).
Furthermore we shall concentrate on so-called {\it alternating}
objects which have much simpler properties than their generic counterparts.
Even though they are probably
asymptotically subdominant (this has been proved in the case of links \STh),
they are quite interesting to study if only because one can prove much more
about them than about general knots. In fact, the number of prime alternating
{\it tangles} is known exactly \STh; however, a general tangle has an arbitrary
number of connected components, and in the present paper we want to be able
to control this number, that is to count knots or similar objects
(e.g.\ tangles with exactly $2$ connected components).

The key concept that will be used is that of a {\it transfer matrix}.
A standard object of statistical mechanics, where it describes the discrete
time evolution of a system, the transfer matrix is also a very efficient
numerical tool for the combinatorial enumeration of discrete objets.
This approach has recently been used to investigate 
the properties of 2D Lorentzian gravity \Lorentz,
the enumeration of plane meanders \refs{\Jensen,\Meanders},
and the coupling of matter fields to 2D quantum gravity equipped
with a Hamiltonian circuit \Ising.
In contradistinction to the standard situation, where the transfer matrix
is used to construct the partition function for a statistical mechanics
system on a semi-infinite strip of finite width, these combinatorial
applications possess a state space which is different in each time slice.
Accordingly the underlying physical models are not defined on a regular
lattice, but belong to the realm of two-dimensional quantum gravity.
In all these examples, the common feature that ensures the existence
of a transfer matrix is that the objects under consideration permit
a time ordering. For the meanders (resp.\ the Hamiltonian circuits) this
was attained by ``reading'' each object as one moves along the river
(resp.\ the circuit), adding one intersection (vertex) at each time step.
Clearly, this stategy in also appropriate for enumerating knots with {\it one}
connected component.

This type of enumeration problems have many interesting connections with 
mathematical physics \refs{\Lorentz{--}\KAUF}. In particular, in 
\refs{\ZJZ,\PZJ,\ZJZb} the counting of links, knots or tangles was reduced
to the evaluation of integrals over $N\times N$ hermitian matrices,
in the limit $N\to\infty$.
Though these integrals cannot be computed explicitly in general, the subject of
matrix models and its well-known connection to 2D quantum gravity \DFGZJ\ provide 
some information
on the universal quantities of the model. This led to conjectures
of the {\it asymptotic} behavior of the number of such objects when the number
of crossings goes to infinity; and one motivation of the present work is to
check the conjecture concerning the asymptotic number of prime alternating knots.

The paper is organized as follows:
after some brief definitions in section 2, which enable us to give a more
precise meaning to the problem that we are addressing,
the basic principles underlying our transfer matrix approach
are described in section 3. 
A number of practical details concerning our
implementation of this algorithm are then given in section 4. Finally,
section 5 gives the results of the numerical enumeration of prime
alternating knots. As a byproduct we also count several
other objects of combinatorial interest.

\newsec{Basic definitions}
A {\it knot} is a smooth circle embedded in $\R^3$, considered up to
homeomorphisms of $\R^3$. In this paper we shall study
slightly different objects, namely knots with ``external legs'';
a knot with $2n$ external legs 
is a a collection of $n$ intervals embedded
in a ball $B$ and whose endpoints are given distinct points
on the boundary $\der B$, considered
up to orientation preserving homeomorphisms of $B$ that reduce to the
identity on $\der B$. These knots with external legs are
nothing but tangles in which no closed loops are allowed.
Knots with $2$ and $4$ external legs
will be of special interest to us. 
When we refer to standard knots we shall from now on call them
closed knots. Clearly, each knot with $2$ external legs 
can be transformed into a closed knot
by joining its endpoints through a smooth curve outside $B$.
And conversely, by cutting a closed knot once it may be turned into a
knot with $2$ external legs,
but in general this transformation is not unique, in the sense
that the topological properties of the resulting knot with $2$ external legs
depend on the point where the closed knot was cut. This means that counting
knots with external legs does not help count closed knots.
However we shall always count the former
and never the latter; those are the objects whose generating series
will have ``nice'' properties.\foot{At the level
of diagrams (see below for a definition of diagrams associated to knots),
one can go slightly further by saying that,
as the Feynman rules of perturbative quantum field theory tell us,
counting planar diagrams with external legs is equivalent
to counting closed planar diagrams weighted by a symmetry factor. This also
explains why counting closed objects with a weight of $1$ is unnatural.}

It is common to represent such objects by their projections on a plane;
we consider regular projections with only double points where lines cross.
To avoid redundancies, we shall concentrate on {\it prime}
knots, whose diagrams cannot be
decomposed as a sum of pieces connected to each other by only two lines, and on 
{\it reduced} diagrams that contain no irrelevant crossings (Fig.~\nprime). 
We shall need another related concept: a diagram is said to be
{\it $r$-particle reducible (resp.\ irreducible)} ($r$PR, resp.\ $r$PI),
$r$ positive integer, if it can (resp.\ cannot) 
be divided into disconnected components by cutting $r$ edges.
For closed knots and knots with $2$ external legs, 
the notion of a prime knot coincides with the
$2$-particle irreducibility of its diagram(s).\foot{Note however
that this is not the case for knots with $2n$ external legs,
$n\ge 2$; see for example
\ZJZ\ for a discussion of 2PI tangles.}
\fig\nprime{a) a non prime link; b) an irrelevant (or ``nugatory'') crossing.}
{\epsfxsize=6cm\epsfbox{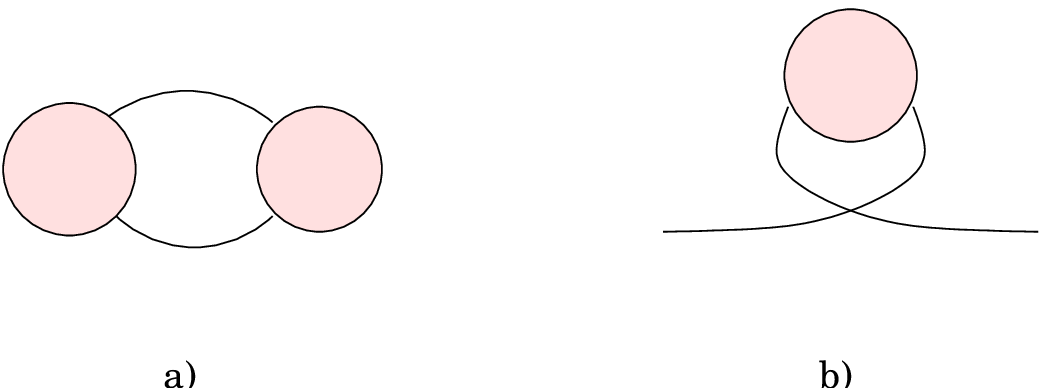}}

A diagram is called {\it alternating} if one meets
under- and over-crossings alternatingly as one travels along each loop. 
A remarkable property (see e.g.\ \KAUF, page 21) is that in the case of
alternating diagrams there is no need to ``remember'' which
crossings are under/over. In other words, two alternating knot diagrams
have the same underlying planar diagram if and only if they are identical,
or related by an overall flip under $\leftrightarrow$ over
in the case of closed knots.
This greatly simplifies their enumeration.

To a given knot can correspond several diagrams. In fact, in the
case of alternating diagrams,
two alternating reduced knot diagrams represent 
the same object if
and only if they are related by a sequence of moves acting on tangles 
called ``flypes'' (see Fig.~\flyp) \MTh. This is of course an essential
distinction when one is interested in counting such objects, and we
shall briefly discuss it now.
The general idea is the same as in \ZJZb; however, the actual
equations shown used here are simpler than those in \ZJZb, and their
proof will be given elsewhere \PZJc\ in the more general framework
of colored links.
\fig\flyp{The flype of a tangle.}{\epsfxsize=6cm\epsfbox{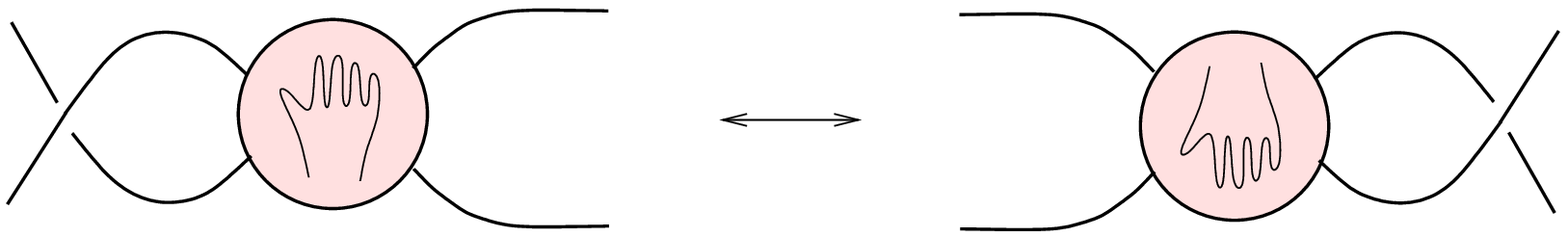}}

Since the diagrams we shall work with most of the time are
diagrams of knots with $2$ external legs, we shall simply call them
knot diagrams.
Let us start by defining the generating series $G(g)$
\eqn\gen{G(g)=\sum_{p=0}^\infty a_p g^p}
where $a_p$ is the number of knot diagrams with $p$ crossings, or equivalently,
the number of topologically inequivalent open curves
in the plane going from $(-\infty,0)$ to $(+\infty,0)$ 
with $p$ regular self-intersections.
We similarly define
$\Sigma_1(g)$ to count 1PI knot diagrams, and $\Sigma_2(g)$ to count 2PI
knot diagrams (with the trivial diagram excluded and the
two diagrams with one crossing included). 
The following relations hold: $\Sigma_1(g)$ is simply given by
\eqn\genb{G(g)={1\over 1-\Sigma_1(g)}}
whereas $\Sigma_2(g)$ is given by the implicit equation
\eqn\genc{1+\Sigma_2(g)=G\left({g\over(1+\Sigma_2(g))^2}\right)}
(see \ZJZ\ for details).

Next, we want to take into account the flyping equivalence in order
to count the actual objects and not diagrams.
The data of $G(g)$
is insufficient for this purpose; we need a more general object,
a double generating series $G(g_1,g_2)$
\eqn\genc{
G(g_1,g_2)=\sum_{p_1,p_2=0}^\infty a_{p_1,p_2} g_1^{p_1} g_2^{p_2}
}
where $a_{p_1,p_2}$ is
the number of topologically inequivalent open curves in the plane 
(going from $(-\infty,0)$ to $(+\infty,0)$)
with $p_1$ regular self-intersections and $p_2$ {\it tangencies}, see
Fig.~\tang.
\fig\tang{Open curve with self-intersections (green dots) and tangencies (red dots).}{\epsfxsize=5.5cm\epsfbox{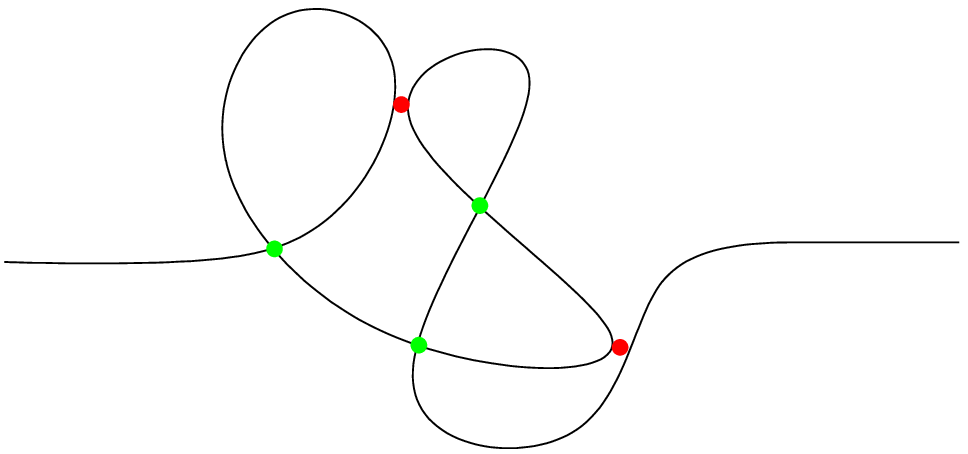}} 
Also, note that we have $a_{p,0}=a_p$ or $G(g)=G(g_1=g,g_2=0)$, so that
there is more information in $G(g_1,g_2)$ than in $G(g)$.

Since flypes act on tangles, we are led to the
introduction of a few more generating functions which are related
to the counting of tangles, $G_1$, $G_2$, $\Gam_1$ and
$\Gam_2$; they are all expressible in terms of $G$ alone via:
\eqna\gend
$$\eqalignno{
G&=1+2g_1G_2+2g_2(G_1+G_2)&\gend a\cr
{\der\over\der g_2} G_2&={\der\over\der g_1} (G_1+G_2)&\gend b\cr
\Gam_1&=G_1&\gend c\cr
\Gam_2&=G_2-G^2&\gend d\cr
}$$

We also introduce for our convenience an extra
parameter $t$ which counts the number of edges of the diagram; it is easy to show
that the following formulae take care of it:
$G(g_1,g_2,t)\equiv{1\over t} G(g_1/t^2,g_2/t^2)$ and $\Gam_i(g_1,g_2,t)\equiv {1\over t^2} \Gam_i(g_1/t^2,g_2/t^2)$.

The parameters $t$, $g_1$ and $g_2$ must then 
be chosen as a function of $g$ according
to the following {\it renormalization procedure} (see \PZJc):
\eqna\ren
$$\eqalignno{
1&=G(g_1(g),g_2(g),t(g))&\ren a\cr
g_1(g)&=g(1-2H'_2(g))&\ren b\cr
g_2(g)&=-g(H'_1(g)+V'_2(g))&\ren c\cr
}
$$
where $H'_1(g)$, $H'_2(g)$ and $V'_2(g)$ are auxiliary quantities defined by:
\eqna\gene
$$\eqalignno{
H'_2\pm H'_1&={(1\mp g)(\Gam_2\pm\Gam_1)\mp g
\over 1+(1\mp g)(\Gam_2\pm\Gam_1)\mp g}&\gene a\cr
V'_2&=(1-g)\Gam_2(1-H'_2-H'_1)^2&\gene b\cr
}$$
where we have omitted all arguments for the sake of brevity;
in particular $\Gam_i\equiv\Gam_i(g_1(g),g_2(g),t(g))$.

This ensures that the flypes are appropriately taken into account and
for example that $\Gam_1(g_1(g),g_2(g),t(g))$ and $\Gam_2(g_1(g),
g_2(g),t(g))$
are the desired generating functions for the number of tangles with
$2$ connected components of type
1 and 2 respectively (see \ZJZb\ for a definition of type; the total
number of tangles is given by $\Gam_1+2\Gam_2$). Similarly one could define
other generating functions of the 2 variables $g_1$ and $g_2$ (higher
correlation functions in the matrix model language)
which would count objects with more external legs.

\newsec{Space of states and transfer matrix}
We now come to the description of the transfer matrix approach. 
The latter requires
first that the knot diagrams be represented in an appropriate way (3.1);
next we 
have to define the space of states on which the matrix acts (3.2); and
finally define the transfer matrix itself (3.3).
At first we shall concentrate
on the usual knot diagrams with self-intersections only; 
a direct application of the first 3 subsections
leads to the enumeration of alternating knot diagrams, but to count actual 
alternating knots, one has to introduce a refined procedure (addition
of tangencies) to which subsection 3.4 is devoted.

\subsec{Representation of a knot diagram using time slices}
A basic ingredient of the transfer matrix approach is the ability
to cut the object one is studying into slices, which
represent the state of the system at fixed (discrete) time.
If we apply this to knots a complication arises.
The naive idea would be to draw the knot diagrams on the plane in such
a way that time would correspond to one particular coordinate of the plane,
that is to read the knot diagrams ``from left to right''. Here,
this idea does not work directly, and one is led to a slightly more
sophisticated notion of slices, which we shall explain using
the example of Fig.~\slice.
\fig\slice{A knot diagram and its corresponding ``sliced'' diagram. The steps
are ordered by their number below the diagram. At
each step, the active line is distinguished by an arrow.}{\epsfxsize=12cm\epsfbox{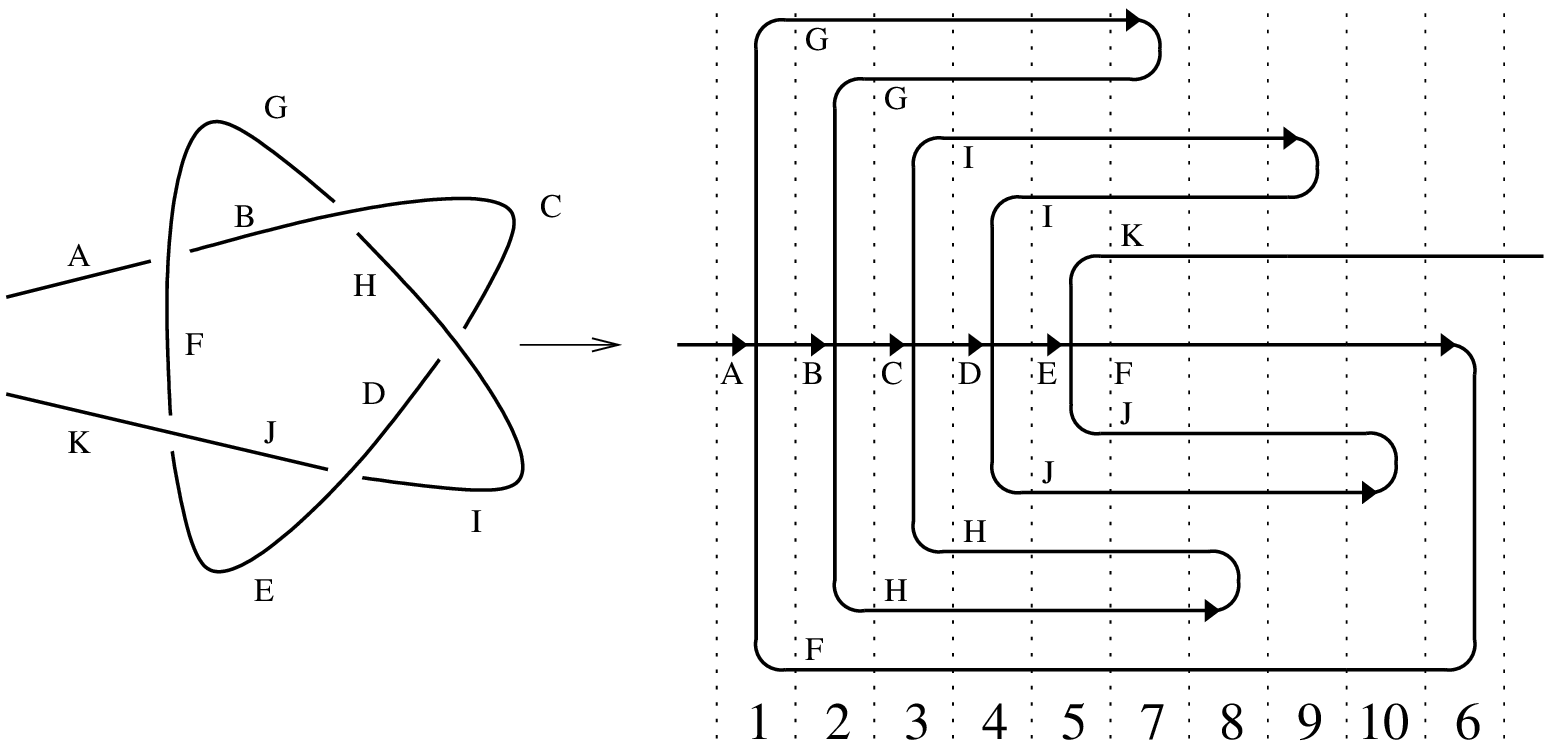}}
The general idea is to follow the knot as it winds around itself from one
``incoming'' external leg to the other ``outgoing'' leg, and write down step
by step the crossings and the lines that are crossed. We shall call the
edge of the diagram that we are currently following the active line.
Of course this is a step-dependent concept since
every edge of the diagram will at some point become the active line.
The edges of the diagram have been labelled from A to K in the order in which 
we encounter them. The precise recipe is as follows.
At each step there are two possibilities:
1) The label of the active line
does not appear anywhere else in the picture drawn this far.
We then proceed to the next crossing and draw it.
2) The active line already appears somewhere. We connect the active line
to its other appearance. We then follow this
new line until it reemerges as an open line: this will be the new active line. 
In the case of Fig.~\slice, steps 1--5 are of type 1) whereas
steps 6--10 are of type 2). In general, steps of type 1) and 2) can
appear in any order, except that at any stage the number of type 2) steps
performed cannot be greater than the number of type 1) steps.
In a more algebraic language, the sequence
of 1s and 2s forms a Dyck word \Dyck.

After step 5, for the first time the active line, which is
now the edge F, is already present (AB--FG crossing). When we reconnect the
two occurrences of the edge F,
we notice that
some open lines are ``imprisoned'' inside the new arch we have created, and
therefore we cannot draw step 6 right after step 5. Instead we must allow
lines inside the new arch to continue to evolve (steps 8 and 10), 
keeping in mind that they
cannot have any contact with the lines outside the arch.\foot{Note that
the order in which we {\it draw} the steps from left to right can contain some
arbitrariness for more complex knots; however the sequence of steps is unique.}

Using this procedure, to each knot diagram we can associate a ``sliced''
diagram; and it is easy to show that two ``sliced'' diagrams are 
topologically equivalent
(in the sense of graphs) if and only if they consist of the same sequence of steps.
We shall now show how to generate all knots with such
diagrams using a transfer matrix.

\subsec{Space of states}
The vector space on which the transfer matrix acts will be spanned
by the intermediate states created
in the process described in the previous section. The important point to remember
is that at each step, we had to take into account the following information:
a) the current (open) lines, including the active line; b) the existing connections
of the lines from the left: pairs of lines are connected by what
we call {\it left arches};
c) the different groups of lines which can still be connected to each other
from the right; the lines
are divided in multiple connected components by what we call {\it right arches}.
All this information has to be included in the state of the system. 

A basis state will therefore be described by a series of left and right arches
and the position of the active line. As an illustration,
we show all the intermediate states of the example of Fig.~\slice\ on Fig.~\states.
\fig\states{A sequence of intermediate states. The active line is denoted by an arrow.}{\epsfxsize=13cm\epsfbox{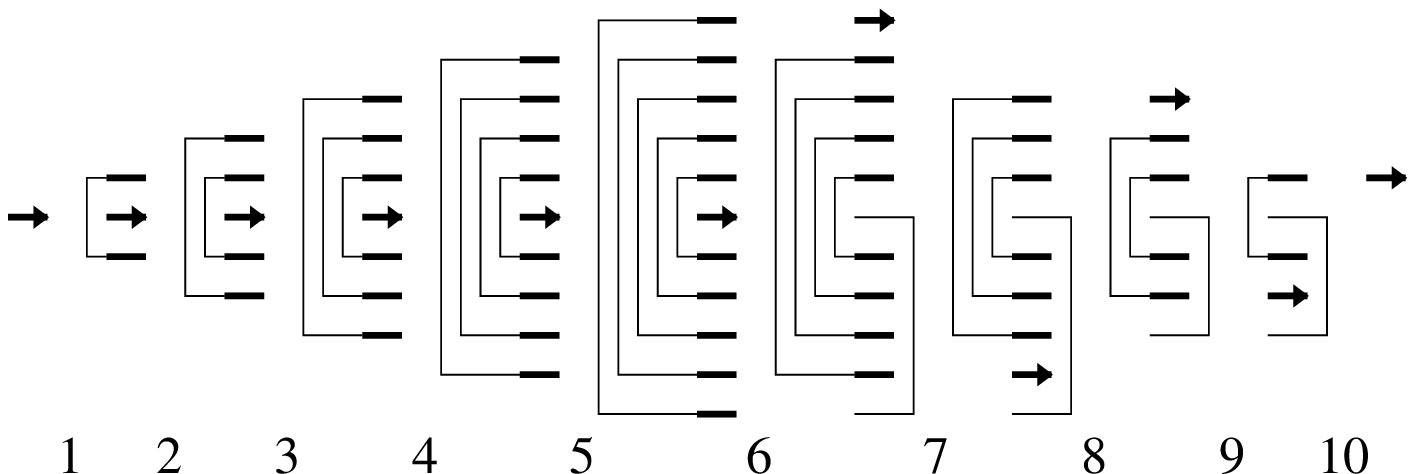}}

For practical applications, it is important to notice that 
some configurations should be forbidden. Firstly, we have states that cannot evolve
into knots (Fig.~\forbid~a) and b)). Secondly, there are redundant states that
are equivalent to simpler states (Fig.~\forbid~c) and d)); we shall
describe a systematic simplification procedure in section 4 below.
\fig\forbid{Examples of forbidden configurations.
a) A region enclosed within a right arch which has no connections with the outside.
b) A region enclosed within a right arch with an odd number of lines.
c) An empty right arch.
d) Two consecutive opening right arches.}{\epsfxsize=6cm\epsfbox{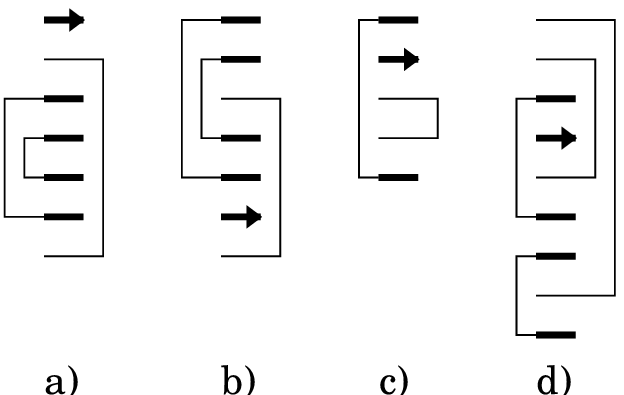}}

\subsec{Transfer matrix}
We now describe the transfer matrix $T$. Its entries $T_{ab}$, where
$a$ and $b$ are two basis states of the kind defined in the previous section,
are $0$ unless $a$ is a descendant of $b$, in which case $T_{ab}$ is the number
of ways $a$ can be obtained from $b$. An allowed state $a$ is a descendant
of $b$ if it can
be obtained via a transformation of one of the two types
shown on Fig.~\transfo, followed by an arbitrary number of simplifications
(Fig.~\simpli).
\fig\transfo{The two types of transformations.}{\epsfxsize=8cm\epsfbox{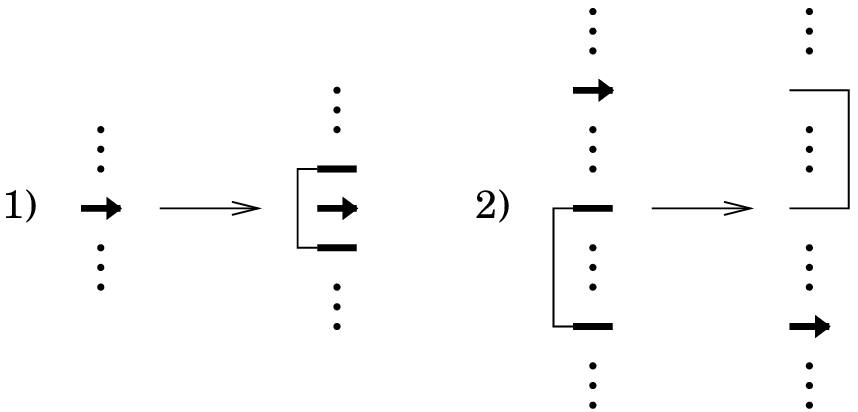}}
\fig\simpli{Possible simplifications.}{\epsfxsize=12.0cm\epsfbox{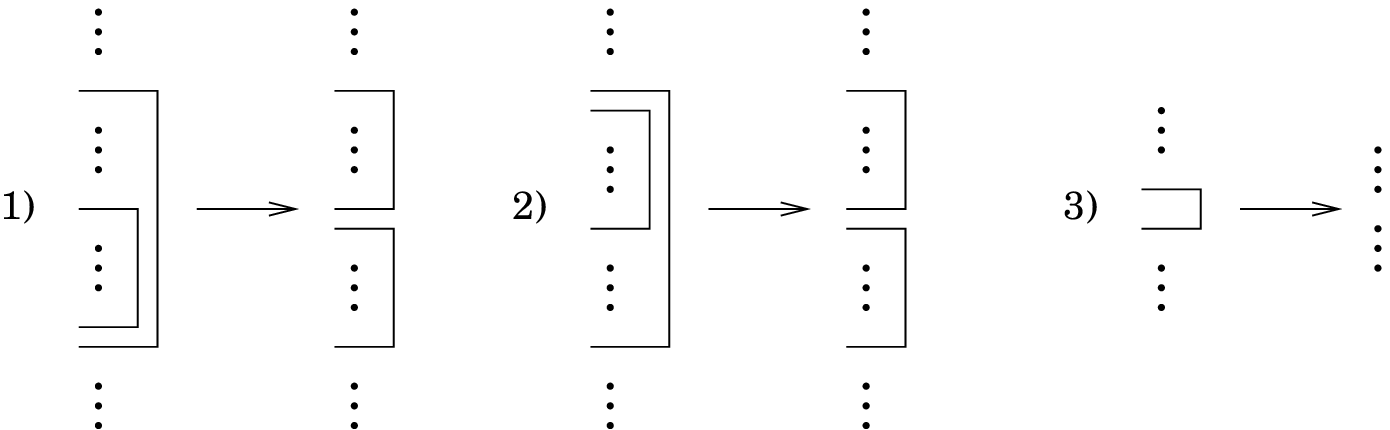}}
The two transformations of Fig.~\transfo\ reproduce the procedure described
in section 3.1, but we have now reformulated it in terms of states.

Note that in transformation 2), the relative
positions of the active line and the line onto which it connects are not
arbitrary. The screening role of the right arches should be
respected, so that the active line can only connect onto lines belonging
to the same block; and the relative distance between the
active line and the line that it connects onto must be {\it odd}.
For each of the allowed connection, the position of the new active line
is found by following the line just connected along the left arch to which
it belongs.

Transformation 1) (resp.\ 2))
increases (resp.\ decreases) the number of lines by 2. This means that
at step $p$, for any intermediate state, 
the number of crossings $n$ and the number of lines $l$ (excluding the
active line) are related by $l=2(p-2n)$.
In particular, we claim that the number of knot
diagrams with $n$ crossings is given by 
$\bra{0}T^{2n}\ket{0}$, where $\ket{0}$ is the state with the active line
only. In this notation, $\ket{0}$ is understood to be assigned the weight
one, and $\bra{0}$ acts as a projection operator: $\braket{0}{0} = 1$.
Formally we have
\eqn\claim{G(g)=\bra{0}{1\over 1-gT^2}\ket{0}}

In Appendix A we show explicitly the action of $T$ in the case of 1PI diagrams
with at most four crossings.

\subsec{Inclusion of tangencies in the transfer matrix}
In order to count knots and not knot diagrams, it was explained in section 2
that one must start with more general objects than standard diagrams: one must
count curves with both self-intersections and tangencies which produce
a double generating series $G(g_1,g_2)$.
This can be easily included in the transfer matrix
as follows.

Firstly, the space of states must be slightly extended to take into account
the fact that we have a double generating series. Typically a state must contain
the information concerning the number of previous tangencies in the diagram.
Therefore the new space of states will be the tensor product of the old space
of states and of the space of polynomials in a variable $x$ which can be
defined as $x=g_2/g_1$.

Secondly, the transfer matrix itself must be modified to allow for the creation
of such tangencies: the new transfer matrix $\tilde{T}$ is of the form
\eqn\newT{
\tilde{T}(x)=T+xT'
}
where $T$ is the old transfer matrix defined by the transformations of 
Fig.~\transfo, and $T'$ is the additional transformation described on
Fig.~\transfob\ (plus, in each case, an arbitrary number of simplifications
of the type of Fig.~\simpli).
\fig\transfob{The transformations that generate tangencies.}{\epsfbox{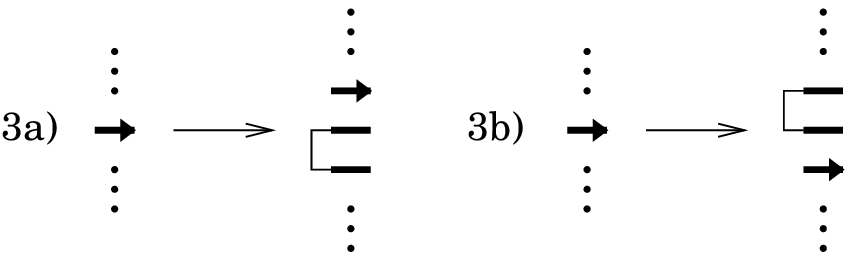}}
We can finally write the following formal expression for $G(g_1,g_2)$:
\eqn\claimb{G(g_1,g_2)=\bra{0}{1\over 1-g_1 \tilde{T}^2(x)}\ket{0}}
where it is recalled that $x=g_2/g_1$.

\newsec{Practical details}
Having described the principles underlying the transfer matrix algorithm
we now turn to a number of important practical details concerning its
implementation. Section 5.1 describes the data structures needed to
encode the states and their corresponding weights. Several remarks on
the algorithm will be made in section 5.2, and in section 5.3 we explain
a number of different implementations that we have made, in the aim of
obtaining a reasonable balance between the time and memory needs of the
algorithm.

\subsec{Data structures}
The state space, formulated in terms of left and right arches and the
active line, has been described in section 3.2. Ideally, in order to
obtain a highly efficient transfer matrix algorithm, one would
like to introduce a {\it ranking} among the states. By this we mean
a bijective mapping from the $N$ different states to the set of
integers $\{1,2,\ldots,N\}$. With a ranking at hand, the integer
representation can be used to label the entries of the transfer matrix,
and the arch representation is then used to produce the descendants
of a given state, as described in section 3.3.
In a previous publication, one of the
authors has shown how to obtain this goal in the case of meanders \Meanders;
however, due to the very complicated interplay between left and
right arches we have not been able to make similar progress in the case
of knots.

Fortunately a simpler, and almost as efficient, alternative is available.
Suppose that to each state $i$ we can assign a unique integer
$k_i \in \Z_+$, and devise a function
$f: \Z_+ \rightarrow \{0,1,2,\ldots,P-1\}$ 
that distributes the set of $k_i$'s more-or-less uniformly on the
set $\{0,1,2,\ldots,P-1\}$. By inserting the states $i$ into an
array of noded lists indexed by $f(k_i)$, we can retrieve a given state
$k$ in a time proportional to the mean length of one of the pointer lists,
$t \propto N/P$. This is a standard technique known as {\it hashing} \Hash;
the integer $k_i$ and the function $f$ are known respectively as
the hash key and the hash function.

In the case at hand, a key $k_i$ can be defined by representing each
arch state as a base-four number. Specifically, we read a configuration of
left and right arches from top to bottom, associating the digit 1 (resp.~0)
with the opening (resp.~closing) of a left arch, and 3 (resp.~2) with the
opening (resp.~closing) of a right arch. The 1s and the 0s (resp.~the
3s and the 2s) thus form two interlaced Dyck words \Dyck.
For the computation at order
$p$ crossings there are at most $p$ arches, and the resulting key
is at most $4^p$; we also need a few extra bits to
specify the position of the active line.
The hash function is simply $f(k) = k {\rm \ mod \ } P$, where $P$ is a
large prime which we choose such that $N/P \sim 10$.

For each state in the hash table, we store its key and its weight.
The weight is an integer, but since the number of knots with $p$
intersections grows exponentially with $p$ the weights of the largest
knots considered in this work cause overflow in a standard 32-bit
integer arithmetic. Instead of wasting memory storing double-precision
integers, we took advantage of modular arithmetic \Knuth. This means that
the largest computations were done modulo various coprime numbers
(typically $2^{32}$ and $2^{32}-1$), and the full result was retrieved
from the Chinese remainder theorem.

\subsec{Algorithmic details}
It is possible to perform a number of reductions on the state space.
Although these do not affect the correctness of the algorithm, they
are nevertheless important to implement since they reduce the number
of intermediate states needed in the transfer process, and thus enables
us to go to larger system sizes.

After each of the two transformations shown on Fig.~6, the resulting
arch state can be simplified using the reductions given in Fig.~7.
The idea is to associate each inequivalent ``screened'' state
with a unique configuration of right arches. Reductions 1) and 2)
consists in sliding an exterior right arch over an adjacent interior
arch, and 3) consists in removing right arches that do not screen
any ingoing left line. In the algorithm, these simplifications are performed
recursively until no further reduction is possible. The resulting state
is then unique.

It may happen that after reduction a state is forbidden in the sense
of Fig.~5 a) or b). A first example of this occurs at order $p=4$, and
is shown in Appendix A. Before inserting a descendant state in the
hash table we therefore examine whether each right arch contains
an even number of lines (including the active line), of which at least
one is connected to the exterior.

A final algorithmic detail concerns the possibility of removing tadpole
insertions in the knot diagrams. A tadpole is generated if and only if
a type 1) transformation is immediately followed by a type 2) transformation
in which the active line connects onto an {\it adjacent} line (see Fig.~6).
We can therefore forbid tadpoles if each state encompasses an extra sign
signalling whether the previous transformation was of type 1).
Superficially this would appear to double the number of states needed,
but in fact this is not so, since an important number of states are
only produced when tadpoles are allowed. In practice we found that
the number of signed states without tadpoles, and the number of states
with tadpoles only differ by a few percent. Of course, eliminating tadpoles
directly in the algorithm carries no intrinsic interest, since it is a
trivial matter to do so afterwards by manipulating the generating functions.
However, since the number of tadpole diagrams is,
very roughly, found to be the square of the number of diagrams without
tadpoles, including tadpoles would mean that we would have to carry out twice
as many runs in order to retrieve the full result from the Chinese
remainder theorem. For this reason we opted for the algorithm without
tadpoles.

\subsec{Implementations}
Even though our transfer matrix method is much more efficient than
a direct enumeration of the knot diagrams, it suffers from the drawback
that the dimension of the state space, and thus the memory needs,
grow exponentially with $p$. For a fixed size $p$, the evolution of the
memory dynamically allocated by the hash table as a function of the
discrete ``time'' steps is shown in Fig.~\memory. Near the beginning the
number of states grows exponentially, reaches a maximum after roughly
$3p/2$ steps, and then decreases exponentially towards the end.
For practical reasons we only had about one gigabyte of memory
available for our computations, and this turns out to be a more severe
limitation for the obtainable system size than the computation time available.
We have therefore experimented with several different implementations
that limit the memory needs at the expense of using more time.

\fig\memory{The evolution of the number of intermediate states, as a function
of `time'. The different curves pertain to the enumeration of diagrams with
$p=8,10,12,14$ and $16$ crossings, no tangencies, and tadpoles included.}
{\epsfxsize=10.0cm\epsfbox{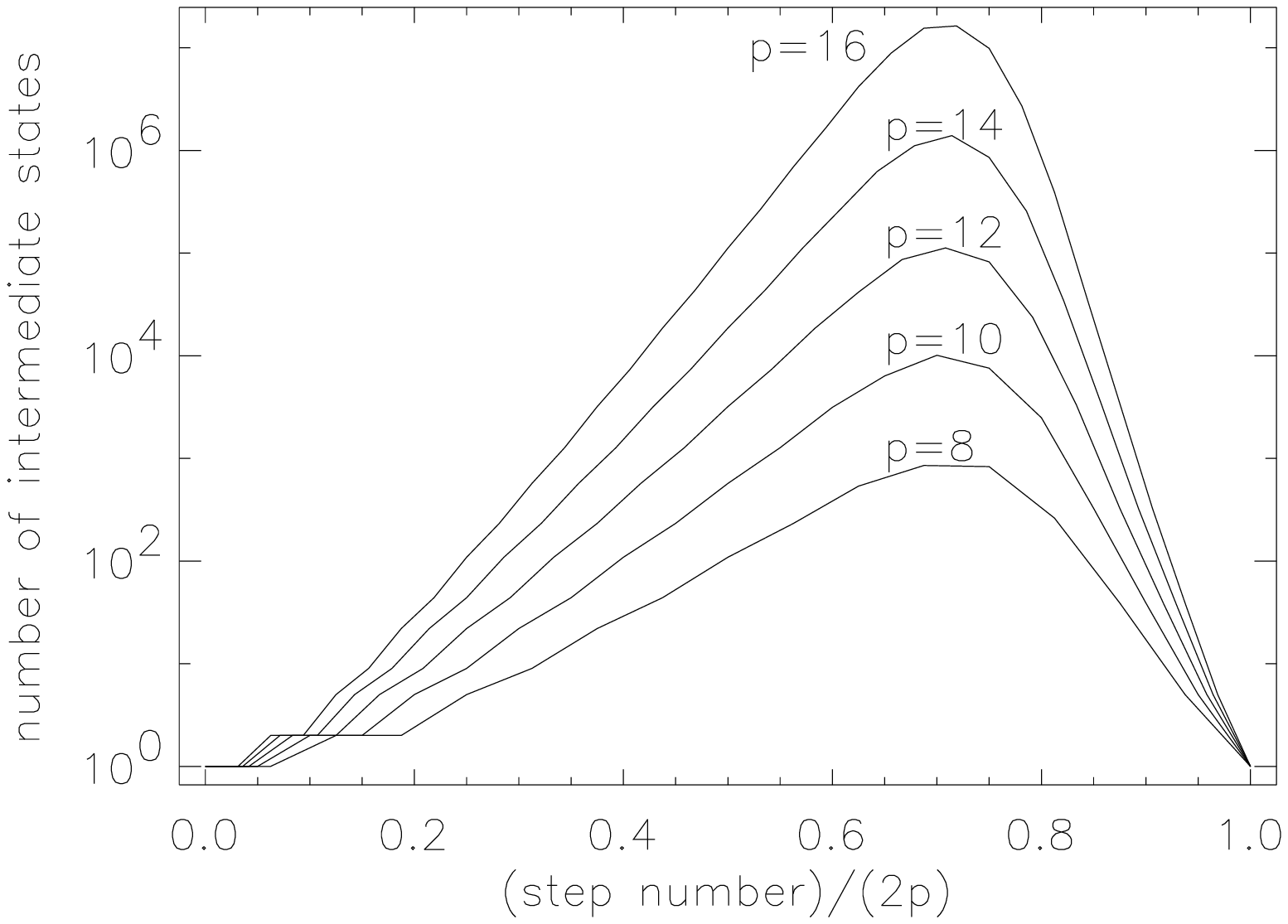}}

The most successful of these consist in, roughly speaking, using the transfer
matrix approach until the available memory is exhausted. We then switch
to a direct enumeration, which is carried out for a fixed number of steps,
until the number of states needed by the transfer matrix approach has
decreased to a level that again fits into the available memory.
On Fig.~\memory\ this could be represented by cutting the maximum of the memory
profile by a horizontal line segment, representing the process of direct
enumeration. The latter is based on the same recursive principle as the
one defining the transfer process, but since states generated in intermediate
time steps are not inserted into the hash table the process allocates
no further memory.

\newsec{Results}
In Table \resdiag\ we display the coefficients of the generating functions
($G$, $\Sigma_1$, $\Sigma_2$) up to order $p=22$.
We recall that these functions represent, respectively, the total number
of knot diagrams with $p$ crossings and two outgoing strings,
and the subsets of $1$PI and $2$PI diagrams.

\tab\resdiag{Table of the total number of two-legged diagrams, as well
as the subsets of $1$PI and $2$PI diagrams.}{\vbox{\offinterlineskip
\halign{\strut\hfil$#$\hfil\quad&\vrule#&&\quad\hfil$#$\hfil\crcr
p&&G&\Sigma_1&\Sigma_2\cr
\omit&height2pt\cr
\noalign{\hrule}
\omit&height2pt\cr
0&&1&0&0\cr
1&&2&2&2\cr
2&&8&4&0\cr
3&&42&18&2\cr
4&&260&108&4\cr
5&&1796&748&12\cr
6&&13396&5648&60\cr
7&&105706&45234&226\cr
8&&870772&378300&1076\cr
9&&7420836&3271204&5156\cr
10&&65004584&29049824&24984\cr
11&&582521748&263656356&128548\cr
12&&5320936416&2436827328&663040\cr
13&&49402687392&22871937208&3514968\cr
14&&465189744448&217536523260&18918792\cr
15&&4434492302426&2092958991474&103123906\cr
16&&42731740126228&20341256951692&569877652\cr
17&&415736458808868&199471121367508&3180066004\cr
18&&4079436831493480&1971730006936240&17921451960\cr
19&&40338413922226212&19630246152650228&101842206548\cr
20&&401652846850965808&196703992506546352&583109887600\cr
21&&4024556509468827432&1982670344984596872&3361640932872\cr
22&&40558226664529024000&20091545428174220376&19499226668816\cr
}}}

The first $10$ terms of $G$ have already been reported by Gusein-Zade
and Duzhin \gusein, who called the corresponding diagrams
`long curves'. The algorithm used by these authors was however based
on direct enumeration, and thus did not enjoy the advantages of the
transfer matrix approach. Namely, in the latter, a multitude of
diagrams can correspond to the same intermediate state at a given time
step, and is therefore counted ``simultaneously''. The difference between
the two approaches can readily be appreciated by comparing the number of
diagrams (Table \resdiag) with the number of intermediate states
(Figure \memory).

In particular, having available more terms of the generating functions
enables us to examine the asymptotic behavior of the number of diagrams.
Calling $a_p$ the coefficients of $G(g)$, as in Eq.~\gen, 
a first rough estimate yields $a_p \sim \mu^p$ with $\mu \simeq 11.4\pm 0.1$.
This corresponds to a singularity of the function $G(g)$ at $g_c=1/\mu$.
However, from the point of view of the underlying field theory it is the
subleading corrections to the dominant exponential behavior that are of
paramount interest, the connective constant $\mu$ being non-universal.
Based on a random-matrix model description of alternating knots
as the $n\to 0$ limit of a generalized $O(n)$ symmetric action
\refs{\ZJZ,\PZJ,\ZJZb}, one of us has conjectured
that the detailed asymptotic behavior reads
\eqn\fitgen{
a_{p} \approx \mu^p p^{-\alpha},}
with $\alpha=3$. This value of $\alpha$ corresponds to 
a string susceptibility exponent
$\Gamma \equiv 2-\alpha = -1$ characteristic of the coupling of a
conformal field theory with central charge $c=-2$ to two-dimensional
quantum gravity, via the celebrated KPZ formula \KPZ.

With the current data, it is difficult to estimate $\alpha$ in Eq.~\fitgen\ 
without any knowledge of the subleading corrections. Indeed, a direct fit gives
$\alpha\approx 2.76$ but usual convergence acceleration methods do not
confirm this result.
However, if we fit our data with $a_p=\mu^p p^{-\alpha} (a\log p+b+o(1))$
(the presence of logarithmic corrections being justified by possible
marginally irrelevant operators in a $c=-2$ theory),
the result is in good agreement with the conjecture:
\eqn\fitgen{
\mu=11.416\pm 0.005 \quad
\alpha=2.97\pm 0.06 \quad
a=0.04\pm0.02 \quad
b=0.1\pm0.03}

In the case of $\Sigma_2$, the approach to the asymptotic regime appears
to be less regular. This is probably due to another singularity
of the function $G(g)$ around $g_{c2}\approx -0.3$ which causes
oscillations that are very subdominant in $G(g)$ but less so
in $\Sigma_2(g)$.
However, the leading exponential behavior
$\sim \mu_2^p$ can be readily extracted without analyzing
numerically the coefficients of $\Sigma_2$, by using the following simple
identity: $\mu_2=\mu/G(g_c)^2$. Assuming the expansion above and using
the fit \fitgen, We find:
$$\mu_2 = 6.613\pm 0.008$$
%For the exactly solvable cases of alternating links with one ($m=1$)
%or two ($m=2$) colors, the corresponding values of $\mu_2$ read
%\refs{\ZJZ,\PZJ,\ZJZb}
%$$ \mu_2(m=1) = {27 \over 4} = 6.75 \ \ \ \
%   \mu_2(m=2) = {16 \over \pi(\pi-4)^2} = 6.91167 \ldots$$
%We can also extract easily the $m\to\infty$ limit \JZJ; we find
%$$\mu_2(m)\buildrel m\to\infty\over\sim 16\, m^{1/2}$$
%We expect
%$\log\mu_2(m)$ to be an increasing concave function of $m$.

%\def\smskip{\hskip.15em\relax}
%\def\sstrut{\hbox{\vrule height5pt depth3.5pt width0pt}}
%\tab\restang{Table of the number of two-legged diagrams with $p_1$ self-intersections and $p_2$ tangencies.}{\vbox{\offinterlineskip
%\halign{\sstrut\hfil$\scriptstyle#$\hfil\enskip&\vrule#%
%&&\smskip\hfil$\scriptstyle#$\hfil\crcr
%{\textstyle{}_{p_1}{}^{p_2}}&&0&1&2&3&4&5&6&\cr
%\omit&height2pt\cr
%\noalign{\hrule}
%\omit&height2pt\cr
\font\five=cmr5
\def\sstrut{\hbox{\vrule height6pt depth3.5pt width0pt}}
\tab\restang{Table of the number of two-legged diagrams with $p_1$ self-intersections and $p_2$ tangencies.}{\vbox{\offinterlineskip
\halign{\sstrut{\five#}&\enskip\vrule#\enskip%
&&\hfil{\five#\ }\hfil\crcr
${}_{p_1}{}^{p_2}$&&0&1&2&3&4&5&6\cr
\omit&height2pt\cr
\noalign{\hrule}
\omit&height2pt\cr
0&&1&2&10&70&588&5544&56628\cr
1&&2&20&210&2352&27720&339768&4294290\cr
2&&8&174&2992&47820&742296&11376554&173401952\cr
3&&42&1504&37100&784672&15294006&283730240&5095814988\cr
4&&260&13300&433620&11515714&271846056&5947557516&123429078160\cr
5&&1796&120744&4928798&158295072&4403552940&111289501120&2626033507768\cr
6&&13396&1122198&55237824&2086803540&66981001600&1923315870960&50921564862176\cr
7&&105706&10638464&614451348&26737722400&973914284112&31351179461568&921163652161792\cr
8&&870772&102541428&6807871480&335676172480&13691869089168&488718870505840&$\ldots$\cr
9&&7420836&1002305040&75275707584&4150940757440&187548130544528&$\ldots$\cr
10&&65004584&9914663308&831595048320&50739269522864&$\ldots$\cr
11&&582521748&99085515840&9185000522880&614607881444256&$\ldots$\cr
12&&5320936416&999104604784&101470031154352&7390867767651290&$\ldots$\cr
13&&49402687392&10153152363648&1121497913694390&$\ldots$\cr
14&&465189744448&103892246982390&12403035430713344&$\ldots$\cr
15&&4434492302426&1069610792999424&137266650274351716&$\ldots$\cr
16&&42731740126228&11072575568623300&$\ldots$\cr
17&&415736458808868&115189593628215600&$\ldots$\cr
18&&4079436831493480&1203690675390892316&$\ldots$\cr
19&&40338413922226212&$\ldots$\cr
20&&401652846850965808&$\ldots$\cr
21&&4024556509468827432&$\ldots$\cr
22&&40558226664529024000&$\ldots$\cr
}}}

We now turn to the inclusion of the flype equivalence.
As described in Section 3.4 this can be done by enumerating also diagrams
with tangencies. A power-counting argument reveals that in order to
accomodate the flype equivalence at order $p$, we need to know the number
of diagrams at order $p_1$ with (roughly) at most
$(p_2)_{\rm max} \equiv \lfloor (p-p_1)/3 \rfloor$ tangencies,
for all $p_1=0,1,\ldots,p$. These data are shown up to order $p=20$
in Table~\restang.

The contents of the first column ($p_2=0$) is of course just the
coefficients of $G$, cf.~Table~\resdiag. The first line ($p_1=0$)
gives the number of two-legged diagrams with $p_2$ tangencies and
no self-intersections,
\eqn\tangonly{
a_{0,p}={(2p)!(2p+2)!\over p! (p+1)!^2 (p+2)!}}
This formula is a corollary of the exact solution of the
standard O$(m\to 0)$ model on random tetravalent graphs. It can also
be shown in a straightforward way.
First, represent each tangency by a dotted line, as in Figure~\tutte.a.
In this way, the problem becomes that of rooted Hamiltonian circuits
on a random trivalent graphs \TutteHam.
Next, straighten out the full line, as in Figure~\tutte.b.
The dotted lines now form two independent Catalan arch configurations,
one above and one below the full line. Clearly, the number of such
configurations is
$$\sum_{j=0}^p {2p \choose 2j} c_{p-j} c_j$$
where $c_k = {(2k)! \over k!(k+1)!}$ are the Catalan numbers.
The result \tangonly\ follows immediately.

\fig\tutte{Counting of pure tangency diagrams. a) Transform each
tangency into a pair of trivalent vertices. b) Stretch the full line.}
{\epsfxsize=10.0cm\epsfbox{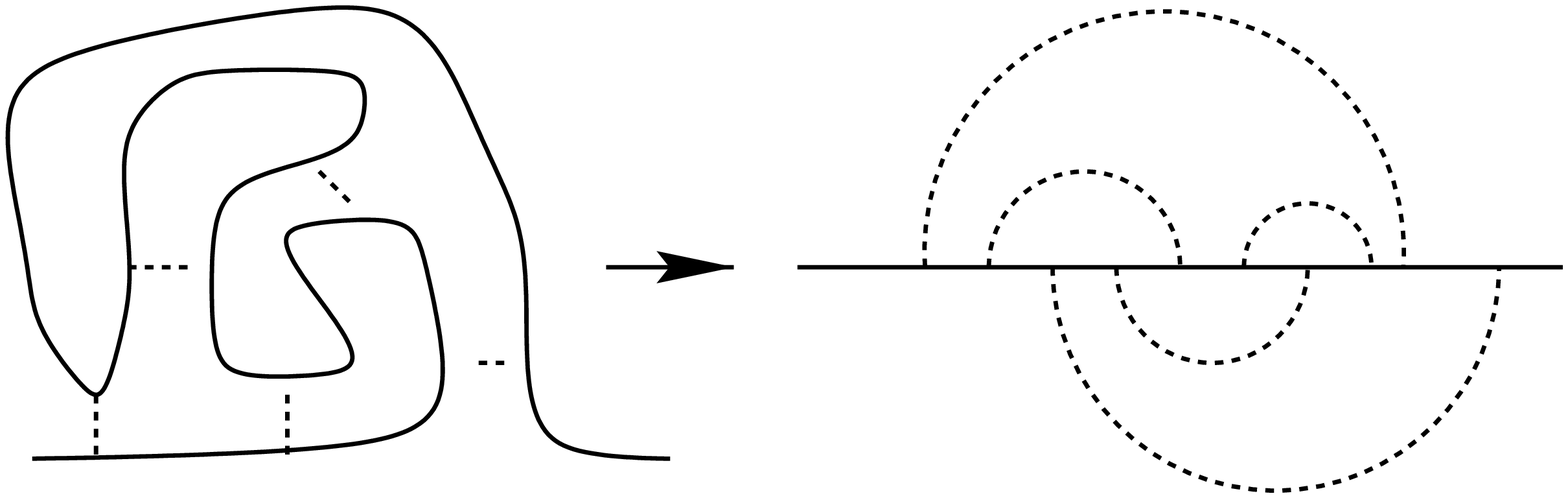}}

In a similar fashion, we can give an explicit formula for the second line:
$$a_{1,p-1}=p \, a_{0,p}={(2p)!(2p+2)!\over (p-1)! (p+1)!^2 (p+2)!}$$
Indeed, the diagrams with one self-intersection and $p-1$ tangencies are obtained
from diagrams with $p$ tangencies by replacing one tangency with a crossing.

\tab\resknot{Table of the number of prime alternating tangles with two
connected components.}{\vbox{\offinterlineskip
\halign{\strut\hfil$#$\hfil\quad&\vrule#&&\quad\hfil$#$\hfil\crcr
p&&\Gam_1&\Gam_2\cr
\omit&height2pt\cr
\noalign{\hrule}
\omit&height2pt\cr
1&&1&0\cr
2&&0&1\cr
3&&2&1\cr
4&&2&3\cr
5&&6&9\cr
6&&30&21\cr
7&&62&101\cr
8&&382&346\cr
9&&1338&1576\cr
10&&6216&7040\cr
11&&29656&31556\cr
12&&131316&153916\cr
13&&669138&724758\cr
14&&3156172&3610768\cr
15&&16032652&17853814\cr
16&&80104192&90220450\cr
17&&408448012&460221672\cr
18&&2105616701 &2365627740\cr
19&&\ldots&12300901598\cr
}}}

The results for the the number of prime alternating tangles with two
connected components can now be found from the procedure outlined in
Section~2; see Table~\resknot. The reader is reminded that the total
number of tangles is given by $\Gam_1+2\Gam_2$.

The first 8 orders were previously
given in \ZJZb. The number of tangles seem to approach their asymptotic
behavior in a less regular fashion than the objects discussed above.
In particular there seems to be a strong dependance on the parity of $p$.
Even the leading term in the large $p$ limit,
of the form $\tilde{\mu}_2^p$, is hard to isolate,
with $\tilde{\mu}_2$ roughly given by $6.0\pm 0.1$; some serious
numerical analysis is required to determine it more accurately.
Note that this
leading behavior should be the same for prime alternating knots, with
the same constant $\tilde{\mu}_2$. Similary, the critical
exponent of knots should be one plus the exponent of tangles.
Although we have not been able to extract reliable values of the
exponent, physical insight suggest that these exponents are most
likely to the same as
those of diagrams (conjecturally, $\alpha=3$).
It is therefore very likely that the conjecture made in \PZJ\ is valid,
though we have no definite evidence at the moment.

In a future publication \JZJ\ we shall show how to generalize our transfer
matrix approach to allow for the enumeration of connected knot
diagrams with an {\it arbitrary} fixed number of components. As a first
application, this will allow us to enumerate alternating {\it links},
and to extend the generating functions given in \ZJZb\ by several
orders. Another interesting goal
that we are currently pursuing is the enumeration of multi-component
meander diagrams \Meanders. There are many other applications related
to the possibility of counting planar Feynman diagrams.
\vfill\eject

\appendix{A}{Knot diagrams up to 4 crossings.}
As an illustration we show the first $8$ iterations of the transfer matrix.
We restrict ourselves to states which generate 1PI diagrams
and with at most $4$ crossings.
Reading off the weight of the vacuum state (containing only the active line)
after step $2p$, we deduce the number of diagrams with $p$ crossings.
However, this state is not allowed to evolve in subsequent steps, since
otherwise one-particle reducible diagrams would be generated.
We have {\it not}\/ excluded tadpoles in this example.

Note also that when connecting the active line of the third diagram at step 6 to the
uppermost line by means of a right arch, we generate (after reduction
of the two right arches) a forbidden
state of the type shown in Fig.~5 a), which is therefore not shown.

\fig\ex{The list of all intermediate steps for 1PI knot diagrams
up to $4$ crossings. The number in subscript is the weight of
the state.}{\epsfxsize=14cm\epsfbox{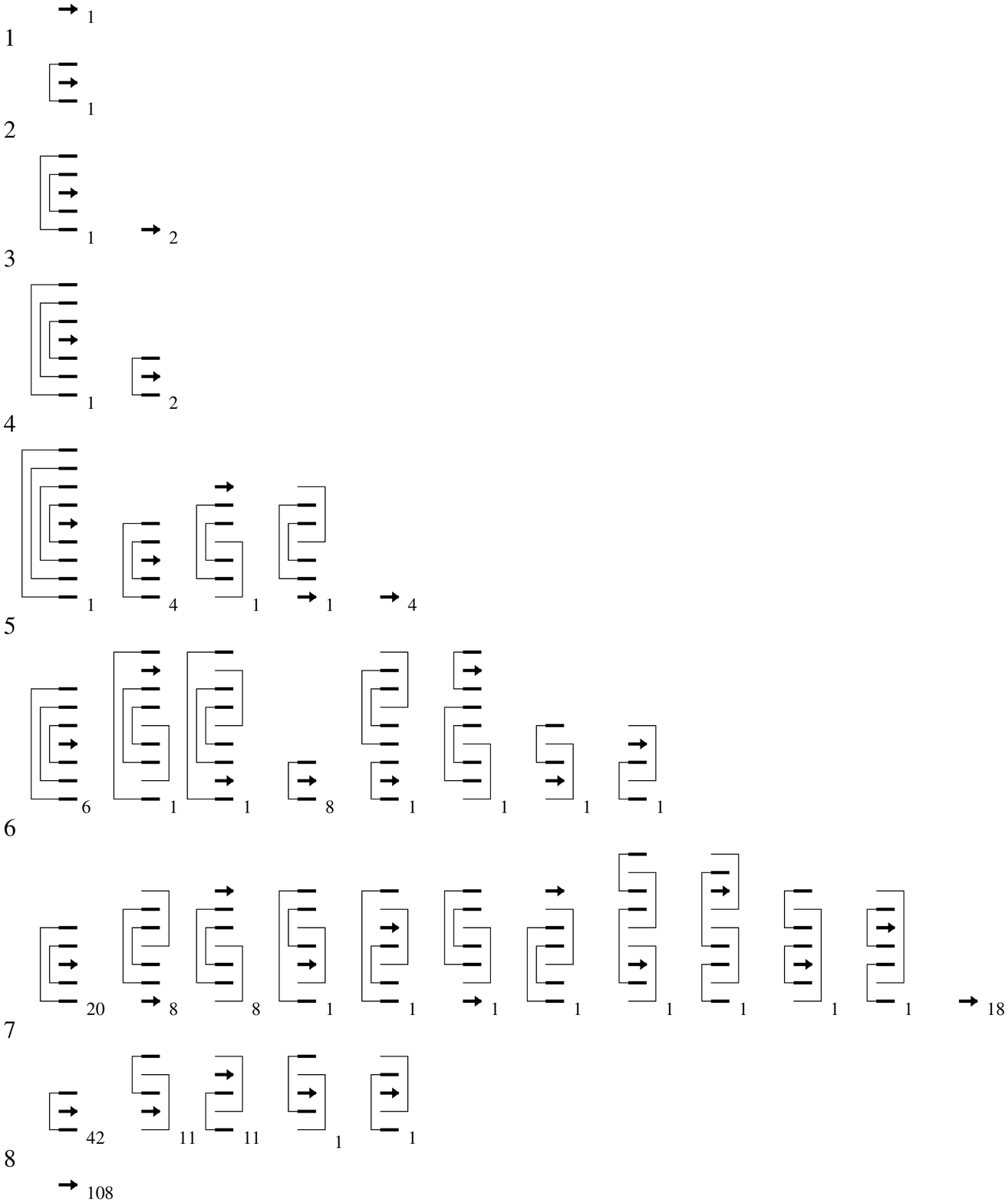}}
\listrefs
\bye